\documentclass[%
 reprint, 
 amsmath,amssymb,
]{revtex4-2}

\usepackage{graphicx}
\usepackage{dcolumn}
\usepackage{bm}

\usepackage[final]{changes} 
\usepackage{caption}
\usepackage[justification=raggedright,singlelinecheck=false]{caption}

\usepackage{makecell}

\usepackage{xr}
\usepackage{xr-hyper}
\usepackage{hyperref}

\usepackage{titlesec}

\titlespacing\section{0pt}{12pt plus 4pt minus 2pt}{0pt plus 2pt minus 2pt}
\titlespacing\subsection{0pt}{12pt plus 4pt minus 2pt}{0pt plus 2pt minus 2pt}
\titlespacing\subsubsection{0pt}{12pt plus 4pt minus 2pt}{0pt plus 2pt minus 2pt}

\makeatletter

\newcommand*{\addFileDependency}[1]{
\typeout{(#1)}
\@addtofilelist{#1}
\IfFileExists{#1}{}{\typeout{No file #1.}}
}\makeatother

\begin{document}

\preprint{APS/123-QED}


\title{Rigid body rotation and chiral reorientation combine in filamentous \emph{E. coli} swimming in low-Re flows}%



\author{Richard Z. DeCurtis}
\affiliation{Department of Bioengineering, Northeastern University, Boston, MA 02115, USA}
\affiliation{Department of Medicine, Rutgers Robert Wood Johnson Medical School, New Brunswick, NJ 08901, USA}
\author{Yongtae Ahn}%
\affiliation{Department of Energy Engineering, Gyeongsang National University, Jinju, Gyeongnam 52828, Republic of Korea}%
\author{Jane E. Hill}
\affiliation{Department of Chemical and Biological Engineering, The University of British Columbia, Vancouver, BC V6T 1Z3, Canada}
\author{Sara M. Hashmi}
\email{s.hashmi@northeastern.edu}
\affiliation{Chemical Engineering, Northeastern University, Boston, MA 02115, USA}
\affiliation{Mechanical and Industrial Engineering, Northeastern University,  Boston, MA 02115, USA}
\affiliation{Chemistry and Chemical Biology, Northeastern University, Boston, MA 02115, USA}


\date{\today}

\begin{abstract}


When treated with doses of antibiotics below the minimum inhibitory concentration, bacterial cell division turns off, but cell growth does not. As a result, rod-like bacteria, including \emph{E. coli}, can elongate many times their original length without increasing their width. The swimming behavior of these filamentous bacteria through small channels may provide insights into how bacteria that survive antibiotic treatment can reach channel walls. Such swimming behaviors in settings like hospital tubing may signal precursors to adhesion, biofilm formation, and infection. Despite the importance of understanding the behavior of bacteria not killed by antibiotics, the hydrodynamics of filamentous bacteria swimming in external flows has not received much attention. We study the swimming behavior of stressed, filamentous \emph{E. coli}.  In quiescence, highly elongated \emph{E. coli} swim with a sinusoidal undulating motion, suggesting rigid body rotation of \deleted{the} long, \deleted{slightly segmented} \added{rigid, buckled} cell bodies.  In low Reynolds number pressure-driven flows through a microchannel, the undulating motion becomes irregular; it may even stop and start within a particular bacterial trajectory.  We refer to this behavior in flow as ``wiggling''.  Rigid body rotation persists in flow, appearing as a high-frequency change in body orientation on top of a slower one. Chiral reorientation can explain the slower reorientation frequency. We quantify swimming behaviors in two different flow rates and observe rheotaxis in addition to the preferential orientation of bacterial bodies. Faster flow constrains wiggling bacteria trajectories and orientations compared to those observed in slower flow, with rheotaxis taking bacteria toward the wall. Interestingly, not all bacteria in flow wiggle. Populations of non-motile “non-wiggling” filamentous E. coli follow streamlines, without preferential alignment of their orientation. Non-motile bacteria do not behave like chiral rods propelled and rotated by flagellar bundles, but like rigid rods. Motility slows the swimmers in comparison. Differentiating these two populations may have important implications for understanding consequences of motility loss that inevitably occurs as bacteria die.

\end{abstract}

\maketitle


\section{\label{sec:intro}Introduction}



When viewed as active colloids, bacteria provide an excellent and tunable model system to investigate a wide variety of phenomena. In addition to providing insights into collective motion and the fundamental physics of systems far from equilibrium\cite{aranson2022bacterial, koch2011collective, zhang2010collective}, the behavior of concentrated bacterial systems drives the growth of biofilms \cite{costerton1999bacterial, donlan2002biofilms}.  Biofilms in contact with implants, medical catheters, or mucus membranes result in various tract infections \cite{sauer_biofilm_2022,lebeaux_vitro_2013}. The clinical importance of biofilms motivates research into the swimming behavior of bacteria in both concentrated and dilute systems \cite{verstraeten2008living, yazdi2012bacterial}.  \added{Bacteria can form clusters while swimming in quiescence through very narrow channels, especially when tumbling becomes frustrated in increasing confinement \cite{carrillo2025preventing}. Exposing bacteria to shear flow can cause their depletion from low shear regions and accumulation in high shear regions, like those found near walls in parabolic flows \cite{rusconi2014bacterial}.} The way in which motile bacteria swim through pores, ducts, and channels determines their ability to attach to walls, a necessary precursor to biofilm formation \cite{figueroa2015living}.  Antibiotics can prevent or remove such infections, but antimicrobial resistance (AMR) is on the rise: discovering new, effective treatments is increasingly difficult \cite{ajulo2024global}.  Additional types of bacteria continue to emerge as highly resistant to antibiotics \cite{ventola_antibiotic_2015}. Drug treatments delivered at concentrations below the minimum inhibitory concentration (MIC) will fail to kill them completely \cite{canton_emergence_2011, tang2023antimicrobial}. 

The surviving bacteria, stressed by the drug treatment, exhibit physiological and metabolic changes which can cause morphological changes.  In particular, the stress of drug treatments can inhibit cell division.  Bacteria continue to grow, but do not divide.  In rod shaped bacteria, this elongation without division is called filamentation \cite{khan2022filamentous, zhang_PNAS2024_filament}.  For instance, \emph{E. coli} are normally short rods, $\sim 2\mu$m long and $<1\mu$m in diameter.  \emph{E. coli} that \deleted{have elongate} \added{elongate} due to stress \added{maintain the location of division sites but do not divide there \cite{wehrens_size_2018}.  As they grow longer, elastic strain builds up in the cell body, which is relieved by buckling \cite{phan2018emergence, nadal2025coli}} \deleted{may appear as multiple connected bacillus rods or as bacteria that appear to be slightly segmented, composed of individual cell bodies that have not fully disconnected from one another.}  Some insights into the swimming \deleted{motion} of filamentous rod-shaped bacteria in external flows can be inferred from the behavior of passive, elongated rods and the behavior of shorter, motile, rod-like bacteria.

Jeffery orbits describe the rotation undergone by passive, non-chiral rods in shear flow \cite{jeffery1922motion, bretherton1962motion}.  Non-active, non-chiral rods exhibit a total rotation about an axis perpendicular to both the direction of flow and the orientation of the rod.  Jeffery orbit frequency increases with both shear rate and the aspect ratio of the rod.  In contrast, a rigid, chiral rod in shear flow undergoes chirality-induced reorientation about a preferred orientation angle \cite{zottl2023asymmetric}. Instead of a constant unidirectional change in orientation angle, chiral reorientation causes rods to continuously oscillate about this preferred angle.  The frequency of chirality induced reorientation scales with both shear rate and the chiral strength of the particle, with an oscillation amplitude that dampens over time \cite{zottl2023asymmetric}.  



``Run and tumble'' swimming behavior of motile, rod-shaped bacteria is observed in quiescent Newtonian fluids.  Motility arises from the rotation of helical flagella. Some bacteria have a single flagellum at one end; rod-shaped \emph{E. coli} have flagella distributed around their entire body \cite{berg.1999}.  Stator motors containing up to 11 units \deleted{each} rotate \added{flagellar filaments counter-clockwise into left-handed helical bundles \cite{lowe1987rapid, wadhwa.2021}.} \deleted{the flagella, which} \added{This} causes counter-rotation by the bacteria cell body  \added{at a rate of $\sim10$ Hz in quiescence \cite{macnab1977bacterial,powers2002role}.  In elongated \emph{E. coli} rigid body rotation ranges from 3-10 Hz \cite{phan2018emergence}.}  Bundling of the rotating flagella causes swimming motion in a single direction, in a ``run''.  Temporary stoppage of the motors compels the flagella to debundle; allowing bacteria to change their swimming direction, in a ``tumble''.  An additional mode of motility in rod-shaped bacteria, a ``tug of oars'' transition between backward and forward swimming, is observed in \emph{B. subtilis} swimming in anisotropic Newtonian media \cite{PRXLife.2.033004}. When swimming in quiescence above horizontal surfaces, bacteria tend to swim in circles counter-clockwise \cite{LAUGA2006400, lauga_bacterial_2016}. \emph{E. coli} swimming in quiescence near horizontal surfaces tend to both stay near the surface and orient the axis of their bodies to point toward it \cite{PhysRevX.7.011010}.  When this swimming is bounded by vertical sidewalls, still in quiescence, \emph{E. coli} tend to swim with the wall on their right \cite{diluzio2005escherichia}. 

When bacteria swim in external flows, the hydrodynamic disturbances caused by the bacteria and their rotating flagella couple with the hydrodynamics of the surrounding environment to cause rheotaxis.  In pressure driven, microchannel flows, these interactions can cause rod-like bacteria to swim toward channel walls and even upstream.  Most experimental evidence comes from studies on motile \emph{E. coli}. With left-handed helical flagella, \emph{E. coli} swim in the vorticity direction, typically downstream and to their left, even if they are swimming in the direction of a wall \cite{marcos2012bacterial, figueroa2015living, mathijssen2019oscillatory}.  \emph{E. coli} swimming upstream can maintain upstream motility until a critical shear rate is reached \cite{hill_hydrodynamic_2007, kaya_direct_2012}. Exceeding this critical point results in advection downstream, again directed toward the sides of the channel \cite{kaya_direct_2012}. \emph{E. coli} can even swim from a reservoir into a channel outlet where fluid is leaving, sometimes swimming macroscopic distances upstream \cite{figueroa-morales_e_2020}. In this case, run-and-tumble motion causes the bacteria to repeatedly localize near channel walls as they swim upstream. A variety of geometrical protrusions anchored on side walls can prevent this upstream swimming, with the goal of preventing infections in catheters \cite{zhou2024ai}. In addition to determining swimming trajectories, rheotaxis can also manifest as chiral reorientation of an asymmetric bacterial rod in shear flow \cite{jing2020chirality}.

Filamentation of swimming rod-shaped bacteria adds a confounding factor to their locomotion, even in dilute systems.  \deleted{However, their swimming motion to date has been analyzed only in quiescent fluids.}  For instance, \emph{Enterobacter} elongated to up to 100$\mu$m exhibit an undulating motion when swimming in quiescence \cite{zhang2022enterobacter}.  This undulation can be explained simply by rigid body rotation of a long, partially segmented rod.  In \emph{E. coli} stressed by sub-MIC concentrations of antibiotics, the resulting filamentous bacteria can become too long to tumble in quiescent media \cite{maki2000motility}. \added{Instead, they tend to simply switch directions between forward and backward motion, with persistence lengths up to 20 times that of normal bacteria\cite{phan2018emergence}.}

Filamentous shape changes coupled with external pressure driven flows are likely to lead to even more complex swimming behavior.  Given the clinical relevance of filamentous bacteria, especially when induced by sub-MIC antibiotics, it is important to quantify and understand their swimming behavior in pressure driven, laminar flows like those found in catheters and hospital tubing.  However, the swimming of very long rod-shaped bacteria in external flows has been largely unexplored.




In this work, we present measurements of filamentous \emph{E. coli} swimming in dilute, pressure driven flows in microfluidic channels. \added{We choose two flow conditions that match shear rates in slow, continuous IV drips in clinical settings.} To stress the \emph{E. coli}, we treat them with the antibiotic cephalexin, below its MIC \cite{maki2000motility}. We find that the resulting filamentous \emph{E. coli} exhibit an oscillatory ``wiggling'' of their elongated rod-shaped bodies in flow\deleted{, with a frequency that increases with the applied flow rate}. Interestingly, the wiggling shape of the swimming bacteria does not follow a sine wave, as observed for rigid body rotation in quiescence. \added{Rather, rigid body rotation couples with a slower frequency chiral reorientation to cause oscillations in the cell body orientation as it swims.  With this coupling, elongated bacteria tend to orient perpendicularly to their own trajectories.} \deleted{Furthermore, we observe trajectories and orientations that depend on flow rate.}  At a lower \deleted{volume flow} \added{shear} rate, \emph{E. coli} trajectories vary widely in their overall direction. At a higher \deleted{flow} \added{shear}  rate, \emph{E. coli} trajectories are more confined to a narrow range of angles, swimming downstream and to their left, toward the wall.  \deleted{Interestingly, the wiggling filamented \emph{E. coli} do not exhibit Jeffery orbits.  Rather, they  tend to orient the length of their bodies perpendicular to their own trajectory directions, especially in the higher flow rate. We can explain the swimming behavior of these stressed bacteria by appealing to a combination of rigid body rotation and chirality induced reorientation.} \added{Together, our results provide insight into the swimming of bacteria stressed, but not killed, by antibiotics in flows through small channels like those relevant in clinical settings.}  




\section{\label{sec:methods}Materials \& Methods}

\subsection{Bacteria Culture}


\emph{E. coli} bacteria, K-12 wild type strain WG1, are grown in tryptone broth (1\% tryptone and 0.5\% NaCl) at 35$^{\circ}$C in a shaking incubator. Cells are cultured for three hours in the presence of sub-MIC cephalexin (20 $\mu$g/mL).  With this duration of treatment, bacterial division turns off.  The \emph{E. coli} elongate and grow to a range of $2.4$ to $10.1 \mu$m long, normally distributed about an average length $\langle l \rangle = 4.9 \pm 1.2 \mu$m, with an average radius $r=0.33\mu$m.  The full distribution is shown in Figure S1.  The bacteria are washed by centrifuging at 5,000g for 5 min, and the pellet resuspended in tryptone broth. The suspension is further diluted to an optical density of 0.04 \added{at a wavelength of 600nm} to ensure optimal cell concentration for tracking individual bacteria.

\subsection{\label{sec:microfluidics}Microfluidic Flow Tests \& Microscopy}


PDMS microfluidic devices are prepared using standard soft lithography methods \cite{duffy1998rapid, mcdonald2000fabrication}.  The device consists of a single channel with a cross section $H = 150 \mu$m wide in the $x$ direction and $H = 150 \mu$m deep in $z$. The \emph{E. coli} suspended in growth media are injected into the microfluidic channel at two different constant volume flow rates, $Q=0.10 \mu$L/min and $0.25 \mu$L/min, using a syringe pump (Harvard Apparatus). The average flow velocity, in $y$, is estimated using $\langle v_f \rangle = Q/H^2 = 74 \mu$m/s for the slower flow and $\langle v_f \rangle = 185 \mu$m/s for the faster flow.  We estimate a wall shear rate using $\dot{\gamma} \sim 2Q/H^3 =$ \deleted{0.5 and 1.2} \added{1.0 and 2.4} 1/s in the slow and fast flows, respectively.  

\added{Our choice of flow rates allows us to match shear rates experienced in slow IV drips.  While faster drips rely on gravity, slow IV drips are accomplished using infusion pumps, for volume flow rates $<10$ mL/hr and as slow as 0.5 mL/hr \cite{skryabina2006disposable, breland2010continuous}.  In 24 gauge tubing with an inner radius  $R=0.265$ mm, wall shear rates estimated using $\dot{\gamma} \sim Q/\pi R^3$ are $O(1)$ 1/s. Volume flow rates $Q\sim O(0.1)$ $\mu$L/min allow us to achieve shear rates of the same order of magnitude, $O(1)$ 1/s, while simultaneously imaging on a microscopic scale in a microchannel.}

In an optical microscope, using phase contrast imaging, we collect videos at 33 frames per second (fps), for 60s for each flow condition.  Spatial resolution is 0.193 $\mu$m per pixel.  Images are 36 $\mu$m wide in the $x$ direction across the channel, and 58 $\mu$m long in the flow direction.  The wall is located at position $x=0$, with flow proceeding in the $y$ direction.  Videos are collected relatively close to the glass coverslip.  A video is also collected of the \emph{E. coli} in quiescent conditions, in growth media in a petri dish. \deleted{Clips of each video are provided in the SI, along with still images showing the definition of the axes.}

\subsection{\label{sec:imageanalysis}Bacteria Trajectories, Orientations \& Shapes}

Using particle tracking and shape analysis, we measure the trajectory of each \emph{E. coli} and the evolution of its orientation and shape in flow.  Images are inverted and binarized, so the individual bacteria appear as bright spots.  In each frame, we identify the centroids of each \deleted{bacteria} \added{bacterium} in $x$ and $y$ and the lengths of its major and minor axes.  Bacteria centroids are fed into standard particle tracking algorithms \cite{crocker1996methods}. We measure instantaneous velocity $v$ in each frame.  We calculate the average velocity of each \deleted{bacteria} \added{bacterial}  trajectory $\langle v \rangle$ using the total distance traveled and the elapsed time $t = t_f-t_0$.  At low $Q=0.10 \mu$L/min, trajectories for bacteria entering the field of view at the top of the frame and leaving \added{the field of view} at the bottom \added{of the frame} have an average residence time \deleted{in the channel} \added{$\tau_r$} $\sim 1.5$s.  When $Q$ is increased to $0.25 \mu$L/min, this decreases to \added{$\tau_r$} $\sim 0.6$s.  

The overall angle of the trajectory, $\beta$, is measured with respect to the flow direction, $y$, and ranges from -90 to 90$^{\circ}$. \emph{E. coli} orientation is also measured in each frame, with respect to the flow direction: $\alpha = 0^{\circ}$ indicates alignment with the flow, and $0^{\circ} \leq \alpha < 180^{\circ}$. The difference between the angle of the bacteria director with respect to its overall trajectory is then given by $Z=\alpha-\beta$. These three angles are quantified in more detail below \added{and illustrated in Figure \ref{DefineAngles}}.

\begin{figure*}
\includegraphics[width=\textwidth]{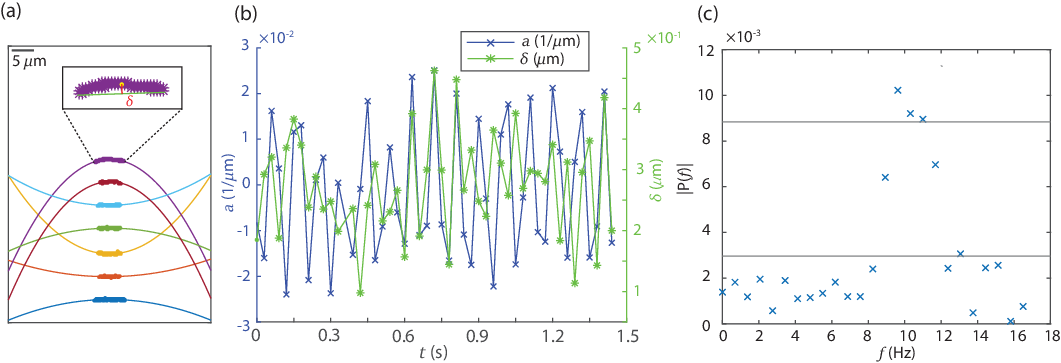}
\caption{\label{DefineWiggle} (a) shows seven \emph{E. coli} isolated from a single image where $Q=0.25$ \added{$\mu$L/min}.  Each has been rotated so its the long axis is horizontal. Solid lines denote parabolic fits to the bacteria shape.  The inset depicts the methodology for calculating the deflection, $\delta$.  (b) shows concavity $a$, in blue, and deflection $\delta$, in green, as a function of time for a single bacteria swimming in $Q=0.25 \mu$L/min.  The lines are to guide the eye. (c) shows the Fourier transform of $a(t)$ shown in (b).  Horizontal lines represent the mean, $\langle |P(f)|\rangle = 0.003$, and two standard deviations above the mean, $|P|=0.009$.  Because the maximum in $|P(f)|$, at $f=9.625$ Hz, is more than two standard deviations above the mean, the maximum frequency is denoted $f_0$ and this bacteria categorized as a wiggler.}
\end{figure*}

To measure the \emph{E. coli} curvature, we fit each shape to a parabola centered at the bacterial centroid using $as^2 + bs + c$, where $s$ is the dimension along the major axis or director.  We use $a$ to characterize the concavity of the bacterial shape.  Analysis of the shape also allows measurement of the instantaneous deflection $\delta$ of the bacteria from the $s$ axis.  \deleted{The} Deflection $\delta$ is calculated by drawing a line connecting the far extremes of each object and measuring the perpendicular distance between this line and the centroid of the object. We track \deleted{these} shape parameters instantaneously for every \added{bacterium} \deleted{bacteria} at every position in its trajectory. Both $a$ and $\delta$ may vary with time through a single trajectory. \added{We analyze the oscillation of $a(t)$ to distinguish motile, `wiggling' bacteria from non-motile, `non-wiggling' bacteria that simply trace the flow. In short, bacteria with a well-defined oscillation frequency of $a(t)$ as determined by Fourier analysis are categorized as wiggling.  Details are provided in Figure \ref{DefineWiggle} and its discussion below.}

\deleted{Where appropriate in the analysis, }\emph{E. coli} populations are distinguished from each other \added{at the two flow rates} by comparing probability distributions \added{of angular orientations, trajectory angles, and oscillation frequencies amongst the wigglers}.  Similarity between populations is assessed by two factor t-tests \added{the F-test of equality of variance}, with $p<0.05$ considered a significant difference between the two populations.  \deleted{We also use the F-test of equality of variance.}

\begin{table*}
    \centering
    \begin{tabular}{|c|c|c|c|} \hline 
         $Q$ ($\mu$L/min) & Category
&  $N$
 & $f_0$ (Hz)
 \\ \hline 
         0.25 & wiggle
&  80&     $9.55 \pm 3.17$\\ \hline 
         0.25 & non-wiggle
&  66&   --
 \\ \hline 
         0.10 & wiggle
&  38&   $7.04 \pm 2.89$\\ \hline 
         0.10 & non-wiggle
&  50&    --
 \\ \hline
    \end{tabular}
    \caption{\deleted{Table 1 indicates} The numbers of tracked bacteria at each flow rate, including both the wigglers and non-wigglers. Wigglers are identified using the analysis depicted in Figure \ref{DefineWiggle}, with average values of $f_0$ reported in the table \added{along with} \deleted{. Error bars represent} the standard deviation.}
    \label{tab:my_label}
\end{table*}


\section{\label{sec:results}Results \& Discussion} 

\subsection{Filamentous \emph{E. coli} wiggle while swimming}




Elongated \emph{E. coli} swimming in quiescence exhibit several interesting features \added{as shown in Figure S2, with a corresponding video online}.  \deleted{The SI contains a video of a collection of \emph{E. coli} swimming in quiescent media (QuiescentRunTumble).}   These bacteria, with typical dimensions, perform the usual run and tumble motion, with velocities $29 \pm 2.4 \mu$m/s, in the expected range of \emph{E. coli} velocities \cite{mears.2014}.  

Several examples are seen in which bacteria are elongated to \deleted{two or three} \added{several} times their normal length.  \deleted{The shapes of these bacteria appear to show segmentation: } Cephalexin prevents complete constriction of the Z ring at sites of cell division, thereby causing elongation or filamentation \cite{lutkenhaus1997bacterial, ebersbach2008novel}.  \added{Eventually the bacteria buckle due to elastic stresses that build up as they elongate, although the cell body remains rigid \cite{phan2018emergence}.} Bacteria which are approximately twice their normal length also perform run and tumble motion similar to untreated bacteria; one of these instances is highlighted in \added{Figure S2 (corresponding video online)} \deleted{the SI video (QuiescentRunTumble)}.  A few longer \emph{E. coli}, $l=4.4$ and $6.9\mu$m, swim persistently in runs without tumbling, with velocities $12.1$ and $11.0\mu$m/s, respectively. \deleted{, both slower than healthy bacteria.}  These each appear to have four or five segments.  An example is highlighted in \added{Figure S3 (corresponding video online).} \deleted{a supplemental video (QuiescentSpotlight).}  \added{ \emph{E. coli} elongated due to stress by cephalexin have been observed to run for up to several mm in length without tumbling at speeds over a range comparable to healthy bacteria \cite{phan2018emergence}.} In these persistently running swimmers, the elongated or filamentous bacteria undulate as they swim. \deleted{in a ``wiggling'' fashion.}  Elongated \textit{Enterobacter} also exhibit undulating motion in quiescence, \deleted{.  In \textit{Enterobacter}, this motion is} explained \deleted{as} \added{by} rigid body rotation \cite{zhang2022enterobacter}. The bacteria rotate around their long axis, and \deleted{a bacteria's shape} \added{the shape of a bacterium} at one time can be rotated and overlaid to match its shape at later times \cite{zhang2022enterobacter}.  


In elongated \emph{E. coli} swimming in quiescence, we demonstrate rigid body rotation in a slightly different manner.  We fit the shape of the bacteria in each frame to a parabola. The parabolic concavity $a(t)$ is well fit by $a(t)=\sin(ft)$.  In the example shown in Figure \deleted{S2} \added{S4}, the oscillation frequency is $f=2.95$ Hz,  \added{consistent with rigid body rotation frequencies in the range of 3-10 Hz observed in stressed \emph{E. coli} \cite{phan2018emergence}.} \deleted{The sinusoidal behavior of $a(t)$ suggests that, in quiescence, elongated \emph{E. coli} swim with an undulating, rigid body rotation.}

\subsection{\added{Defining and quantifying wiggling in external flow}}

Observations of the raw motion of the \added{elongated} \emph{E. coli} in \added{external} flow reveal additional interesting behaviors, \added{as shown in Figure S5, with a corresponding video online} \deleted{as seen in the SI videos (ExampleMethodology)}. The most salient features are the wiggling swimming motion of the bacteria, motion across streamlines or upstream against the flow, and the misalignment of the bacteria director with the flow direction.  \deleted{In what follows,} \added{Unlike in quiescence, wiggling behavior in external flow is not easily described by a sine function.} We first define the wiggling motion in flow before discussing bacteria trajectories and orientations.



Approximately half of the \emph{E. coli} in flow wiggle, with their shape and curvature varying along their trajectories.  That is, the concavity or curvature $a(t)$ oscillates; some \emph{E. coli} curvatures oscillate more than others.  The other half of the population of elongated bacteria exhibit no oscillation in their shape at all, and appear to \deleted{be} \added{behave as} rigid rods as they flow with the background fluid, \added{acting as tracers following the streamlines. This rigid-rod like behavior and absence of rigid body rotation suggests decreased motility.  One possible explanation, beyond the antibiotic treatment, is that exposure to shear can break flagellar filaments, both in \emph{E. coli} and other flagellated species like \emph{B. diazoefficiens} \cite{turner2012growth, carrillo2025damage}.  Depending on the type of filaments damaged, filament regeneration can occur over tens of minutes or several hours. It has been suggested that monitoring motility can be used as a proxy to measure both flagellar filament breakup and regrowth \cite{carrillo2025damage}. A reduction in motility might also signify dead bacteria.}


To differentiate wigglers from non-wigglers in flow, we investigate the oscillations in $a(t)$.  An example of this analysis is shown in Figure \ref{DefineWiggle}.  Figure \ref{DefineWiggle}(a) shows 7 individual bacteria shapes isolated from a single video frame in which $Q=0.25 \mu$L/min; the colorized coordinate points correspond to the bright pixels isolated from the binarized image.  For ease of visualization, each \deleted{bacteria} \added{bacterium} has been rotated so its major axis is horizontal. Deflection from a line, $\delta$, is measured using the distance from the bacteria centroid to the line connecting the two ends, as shown in the inset. Figure \ref{DefineWiggle}b shows both $a(t)$ and $\delta(t)$ for a single bacteria swimming in $Q=0.25 \mu$L/min.  For both $a(t)$, in blue, and $\delta(t)$, in green, the lines connecting the data points are meant to guide the eye.

Interestingly, in contrast to the quiescent example in Figure \deleted{S2} \added{S4}, the dynamics of $a(t)$ for elongated swimmers in flow is generally not sinusoidal.  Two examples of $a(t)$ are shown in Figure \deleted{S3} \added{S6}, with one example showing a reasonable fit to $\sin(ft)$, and the other showing a poor fit.  The situation in which $a(t)$ is not sinusoidal dominates the observations, accounting for more than $95\%$ of wiggling \emph{E. coli}. Therefore, instead of fitting $a(t)$ to $\sin(ft)$, we calculate the Fourier transform of $a(t)$ for each trajectory and analyze the frequency spectrum $|P(f)|$. If the maximum in $|P(f)|$ is more than two standard deviations greater than the mean, we define the location of the maximum as the fundamental frequency $f_0$.  We designate trajectories in this category as belonging to wiggling \emph{E. coli}.  \deleted{Trajectories without well-defined fundamental frequencies in $a(t)$ are non-wigglers.}  The example of $|P(f)|$ in Figure \ref{DefineWiggle}c corresponds to the behavior of $a(t)$ in Figure \ref{DefineWiggle}b. The two horizontal lines in Figure \ref{DefineWiggle}c indicate the mean of $|P(f)|$ and two standard deviations above the mean.  Figure \ref{DefineWiggle}c therefore represents a wiggler: the maximum value of $|P(f)| = 0.01$ is more than two standard deviations above the mean, $\langle |P(f)|\rangle = 0.003$, and occurs at $f_0=9.5$ Hz.  \added{Trajectories without well-defined fundamental frequencies in $a(t)$ are non-wigglers.} Figures \deleted{S4} \added{S7} and \deleted{S5} \added{S8} show two examples of this analysis for non-wigglers.  

Table 1 summarizes the number of \emph{E. coli} in each category at both flow rates, with the mean values of $f_0$.  \added{The average mean values of $f_0$ are $\sim 10$ and $\sim 7$ Hz at $Q=0.25$ and $0.10$ $\mu$L/min, respectively.  The measurements of $f_0$ are well separated from any potential artifacts based on the $\sim1.5$ s duration of the time series (0.67 Hz) or the temporal resolution of the video (33 Hz).} 
\deleted{In the analysis that follows, results are presented mainly for the wiggling \emph{E. coli} at each flow rate, with results for non-wigglers discussed in more detail in the SI.}  
\deleted{The bending behavior $\delta(t)$ seen in the \emph{E. coli} in Figure \ref{DefineWiggle}b} \added{The apparent wiggling} does not suggest the bacteria are flexible. \deleted{, nor is the viscous shear stress of the fluid sufficient to bend them.}  \added{Rather, the elongated bacteria are rigid rods that have buckled due to the build-up of elastic stresses over 10s of minutes during growth \cite{phan2018emergence, nadal2025coli}.  The wiggling of the shape at $f_0\sim10$ Hz in pressure-driven flow reflects rigid body rotation consistent with that observed in quiescence \cite{macnab1977bacterial,powers2002role, phan2018emergence}.}  



\added{We can confirm that the curvature of a bacterium does not arise from bending using known mechanical properties of elongated \emph{E. coli}.} We \deleted{can} assess the possibility of bending in response to shear stress using $\bar{\mu} = 8\pi \mu\dot{\gamma} l^4/B$, a dimensionless number comparing shear stress to bending stress \cite{liu2018morphological}.  Here, $B$ is the bending or flexural rigidity of the rod, $B=YI$ where $I$ is the second moment of inertia and $Y$ the Young's modulus of the cell wall \cite{amir_bending_2014}. Measurements of the bending of elongated \emph{E. coli} grown in a microfluidic ``mother machine'' suggest $Y\sim \text{O}(10)$ MPa and $B\sim \text{O}(10^{-20})$ Nm$^2$ \cite{amir_bending_2014}.  Indentation by atomic force microscopy similarly measures $Y\sim \text{O}(10)$ MPa for the \emph{E. coli} cell wall \cite{deng2011prl}.  Combined with the applied shear rates and measured bacteria lengths in our study, this degree of rigidity suggests $\bar{\mu}\sim10^{-3}$.  This value of $\bar{\mu}$ is orders of magnitude too small to indicate bending: for thin elastic filaments in shear flow, bending occurs beyond $\bar{\mu}\sim10^{3}$ \cite{liu2018morphological}.  

We can also compare bending observed in the mother machine, in response to shear flow, to our measurements of $\delta(t)$ \added{in Figure \ref{DefineWiggle}(a)}.  In the mother machine, one end of \deleted{the bacteria} \added{a bacterium} is immobilized in a micro-well and the other extends into a channel transverse to a pressure driven flow.  When pulses of flow at shear rates $\dot{\gamma}>O(1000)$ 1/s are sent past \added{elongated} \emph{E. coli}, they bend several microns \cite{amir_bending_2014}.  The much gentler $\dot{\gamma}\sim O(1)$ 1/s used in our flow tests would cause \emph{E. coli} to bend with $\delta<1$ nm only.  \added{However, we observe $\delta \sim O(100)$ nm, which is too large to be explained by bending due to shear stress.  Together, these two comparisons} \deleted{Both estimates} suggest \deleted{the bacteria} \added{elongated \emph{E. coli}} are too rigid to bend in our experiments.  \deleted{The curved shape of the \emph{E. coli} more likely suggests that segmentation generates kinks in the elongated backbone.}  \added{Still, we refer to their swimming behavior in external flows as wiggling to differentiate it from the purely rigid-body rotation swimming motion observed in quiescence. In the following analysis, results are presented mainly for  wiggling \emph{E. coli}, with results for non-motile, non-wiggler tracers discussed in more detail in the SI.}






\subsection{\label{sec:trajectories}Faster $Q$ constrains trajectories \added{and} orientation}

\begin{figure}
\includegraphics[scale=0.9]{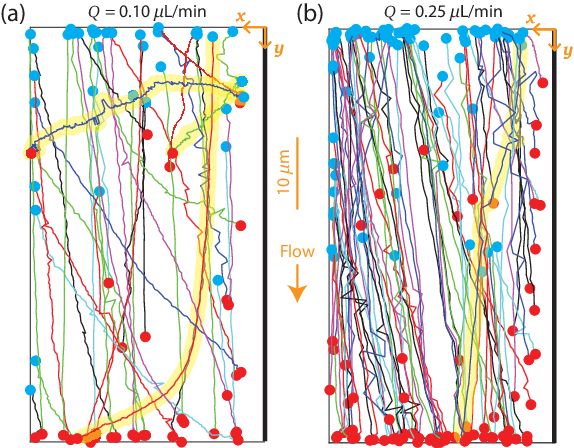}
\caption{\label{Trajectories} Trajectories of \added{38} wiggling bacteria at $Q=0.1 \mu$L/min, in (a), and \added{80 at} $Q=0.25 \mu$L/min, in (b). \added{Trajectories begin at the blue dots and end at the red dots.  Different colors of the solid lines represent different trajectories.} Each image is $36\mu$m in the $x$ direction and $58\mu$m in the $y$ direction.  \added{The upper right corner defines the axes; flow is in the $y$ direction and the scale bar refers to both panels.} The wall is located on the right hand side of each image at $x=0$ \added{indicated by a thicker vertical line.  A few trajectories are highlighted in each panel to show examples of wiggling bacteria that swim away from the wall.}} 
\end{figure}

Figure \ref{Trajectories} shows the trajectories of all wiggling \emph{E. coli} at both flow rates. The axes are labeled with the origin in the upper right: $x=0$ indicates the wall; flow proceeds in $y$. Blue dots indicate the beginning of each trajectory, and red dots the end. Each trajectory is illustrated with a different color line tracing $(x(t), y(t))$ over its length.  Trajectories of $N=42$ wiggling bacteria at $Q=0.10 \mu$L/min are shown in Figure \ref{Trajectories}(a), and $N=79$ wiggling bacteria at $Q=0.25 \mu$L/min in (b).  

The trajectories appear biased toward the high shear-gradient region near the wall. This is especially apparent in the faster flow, which appears to have a narrower range of trajectory angles.  Once the \emph{E. coli} reach near the wall, some turn and swim upstream.  While tracking upstream trajectories is difficult due to the roughness in the PDMS side walls of the channel, a few examples of upstream swimmers are highlighted in \added{Figure S9 (corresponding video online) and Figure S10 (corresponding video online)} \deleted{the SI videos (Upstream)}. Some bacteria appear to swim in place at the wall before falling back into the flow.  Others progress a distance upstream along the wall before falling back into the flow.  While bacteria swim across streamlines at both flow rates, upstream swimming occurs more frequently in the slower flow rate.  

In each panel of Figure \ref{Trajectories}, most \emph{E. coli} trace paths from left to right, in the $-x$ direction toward the wall at $x=0$.  This overall bias in the bacteria trajectories corresponds to the direction of the flow vorticity, $\Omega=-\dot{\gamma}\hat{x}$, and is expected in rheotactic motion  \cite{mathijssen2019oscillatory}. From the perspective of the swimmers, they swim toward the left, from regions of faster flow and lower shear rates toward regions of slower flow and higher shear rates.  This crossing of streamlines is also observed in healthy, non-filamentous \emph{E. coli} \cite{hill_hydrodynamic_2007, kaya_direct_2012}.  Even in the absence of active motion, spheres with rigid chiral tails attached flow across streamlines. When the chiral tails are left-handed, like the flagella of \emph{E. coli}, trajectories are directed to the left, in the $-x$ direction \cite{jing2020chirality, zottl2023asymmetric}.  As far as we know, this prediction of rheotaxis has not yet been reported in experimental investigations of elongated swimmers in external flows, whether the swimmers are stressed \emph{E. coli} or other naturally elongated species. In contrast to wigglers, non-wiggling \emph{E. coli} follow the streamlines of the flow (Figure \deleted{S6} \added{S11}), \added{consistent with the absence of rigid body rotation that indicates a lack of motility}.




\begin{table*}
    \centering
    \begin{tabular}{|c|c|c|||c|c|c|||c|c|c|||c|} \hline 
         $Q$ ($\mu$L/min) & $\dot{\gamma}$ (1/s) & Category
& $\left<\lambda\right>$  ($\mu$m/s)&$\lambda_{p95}$&$\lambda > 1.05 (\%)$ & $\alpha$ ($^{\circ}$) & $\beta$ ($^{\circ}$) & $z$ ($^{\circ}$) &  $\langle v \rangle$ ($\mu$m/s) \\ \hline 
         0.25 & 2.4 & wiggle
&   $1.03$&$1.09$&$22.5\%$  &$95.3 \pm 27.8$&$7.23 \pm 7.20$&$88.0 \pm 27.2$&   $43.4 \pm 22.0$  \\ \hline 
         0.25 & 2.4 & non-wiggle
&  $1.00$&$1.01$&$0.00\%$ & $77.5 \pm 51.3$&$1.41 \pm 6.01$&$76.1 \pm 50.3$&  $93.5 \pm 55.6$\\ \hline 
         0.10 & 1.0 & wiggle
&  $1.05$&$1.13$&$42.4\%$ & $88.0 \pm 50.1$&$6.67 \pm 22.6$&$81.3 \pm 51.5$&   $21.3 \pm 14.6$ \\ \hline 
         0.10 & 1.0 & non-wiggle
&  $1.01$&$1.05$&$5.41\%$ & $84.5 \pm 53.7$&$3.10 \pm 10.8$&$81.4 \pm 54.4$& $38.8 \pm 23.9$ \\ \hline
    \end{tabular}
    \caption{\deleted{Table 2 indicates} Several parameters measured for each population of elongated \emph{E. coli}.  $\lambda$ refers to the trajectory tortuosity defined in Eq. 1, with full distributions of $\lambda$ provided in Figure \deleted{S7} \added{S13}.  The angles $\alpha$, $\beta$ and $Z$ are defined in Figure \ref{DefineAngles}(a), with distributions shown in Figure \ref{AngleHistograms}.  Average values of velocity, $\langle v \rangle$, refer to velocity over the course of an entire trajectory, with distributions shown in Figure \ref{Velocity}.  All error bars, $\pm$, are calculated using standard deviation. }
    \label{tab:my_label}
\end{table*}

\begin{figure*}
\includegraphics{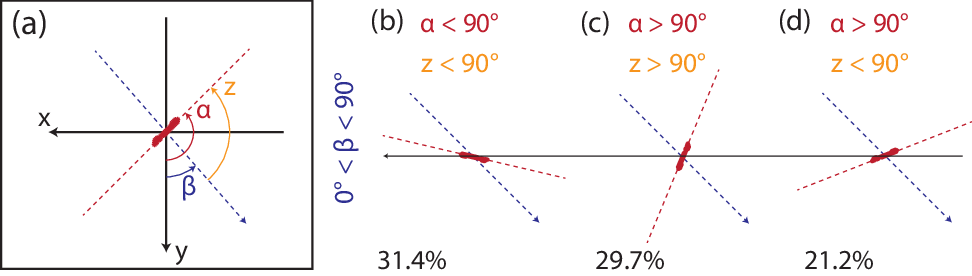}
\caption{\label{DefineAngles} (a) shows the definitions of the three angles used to define the orientation of an \emph{E. coli} and its trajectory.  While the axes are centered on the bacteria for ease of viewing, the location of $x=0$ in flow is at the wall.  Flow proceeds in the $y$ direction.  \deleted{The remaining} Panels \added{(b), (c), and (d)} show examples of the three most common orientations of the average trajectory, $\beta$, average bacteria orientation $\alpha$, and the angle between these, $Z$. The percentages indicate below each panel correspond to the percentage of wiggling \emph{E. coli} at both flow rates that exhibit the behavior pictured.}
\end{figure*}

\deleted{the average angles calculated from the definitions provided in Figure \ref{DefineAngles}.}  

Elongated \emph{E. coli} swimming in slower flow tend to follow more tortuous, less linear trajectories.  In Figure \ref{Trajectories}a, where $Q=0.10 \mu$L/min, multiple \emph{E. coli} travel across streamlines toward the wall along a curved path.  A few trajectories go in the opposite direction, starting from blue dots near the wall and traveling across the flow toward the channel center.  \added{Three examples are highlighted in Figure \ref{Trajectories}(a) and (b), and one is shown in Figure S12, with a corresponding video online.} In Figure \ref{Trajectories}(b), where $Q=0.25 \mu$L/min, most wiggling \emph{E. coli} travel toward the wall with straighter, less curved paths.  

We quantify path tortuosity by a straightness index $\lambda$: 

\begin{equation}
\lambda = \frac{\sum_{t=1}^{t=f} \sqrt{(x_{t}-x_{t-1})^2+(y_{t}-y_{t-1})^2} }{\sqrt{(x_f-x_1)^2+(y_f-y_1)^2}}
\end{equation}

\noindent where the numerator is the total path length of the trajectory and the denominator the shortest distance between its initial ($t=1$) and final ($t=f$) points.  This metric can also estimate tortuosity in animal locomotion \cite{benhamou_how_2004}.  When $\lambda=1$, the trajectory is a straight line; $\lambda\gg1$ indicates a meandering, non-linear path.  


The most tortuous paths are seen in wiggling \emph{E. coli} in slower flow. Table II provides a summary. For the slow and fast flow rates, average tortuosity $\left<\lambda\right>=1.05$ and 1.03, respectively, \added{but the distributions of $\lambda$ are non-Gaussian, as seen in Figure S13.  Therefore we also report} the $95^{\text{th}}$ percentile values \added{to} indicate the tail of the distributions: $\lambda_{p95}=1.13$ and 1.09 in slow and fast flow.  Approximately $42\%$ of wiggling \emph{E. coli} in the slow flow rate have $\lambda>1.05$, thereby exhibiting path lengths at least $\sim5\%$ greater than the minimum.  In faster flow, this percentage drops to $\sim22$\%.  The distributions of $\lambda$ exhibit tails extending as high as $\lambda>1.1$ in the fast flow and $\lambda>1.2$ in the slow flow, as seen in Figure \deleted{S7} \added{S13}.  In contrast to the wigglers, non-wiggling \emph{E. coli} trajectories are straight, with \added{very little if any tortuosity to their paths:} $\left<\lambda\right>=1.01$ and 1.00 in slow and fast flows, respectively.  Also, $\lambda_{p95}=1.05$ and 1.01 in slow and fast flows. In the faster flow, no non-wiggling \emph{E. coli} trajectories have $\lambda>1.014$.  \added{Non-wigglers follow the streamlines, tracing the fluid velocity, again consistent with a reduction in or lack of motility.}

While Figure \ref{Trajectories} shows trajectories, and $\lambda$ quantifies path tortuosity, neither of these metrics captures the \emph{E. coli} orientation.  To do this we quantify three angles: orientation of the \deleted{bacteria} \added{bacterium} director $\alpha$, trajectory direction $\beta$, and the angle between them $Z=\alpha-\beta$, with definitions shown in Figure \ref{DefineAngles}(a) \added{with respect to a bacterium shown in red}. For ease of visualization, the $x$ and $y$ axes are centered on the \deleted{bacteria, shown in red} \added{bacterium in Figure \ref{DefineAngles}(a)}. The red dashed line indicates the \deleted{bacteria} \added{bacterium} orientation angle $\alpha$.  The blue dashed line indicates the \deleted{bacteria} \added{bacterium} trajectory \deleted{defined by} angle $\beta$.  Both $\alpha$ and $\beta$ are measured with respect to the flow direction $y$. When $\alpha=0$, the \deleted{bacteria} \added{bacterium} director is aligned with the streamlines; when $\beta=0$, the trajectory follows the streamlines.  In Figure \ref{DefineAngles}(a), orange \deleted{is used to define} \added{defines} $Z$: when $Z=0$, the \deleted{bacteria} \added{bacterium} director is aligned with its own trajectory\deleted{; that is,} \added{and} the \deleted{bacteria} \added{bacterium} swims ``nose down'' \deleted{or} in the nematic direction.  Swimming with $Z=0$ occurs in quiescence, both in \deleted{the} elongated \emph{E. coli} in this \deleted{study} \added{and other studies}, and in \emph{Enterobacter} \cite{phan2018emergence, zhang2022enterobacter}.  \emph{B. subtilis} also swim persistently with $Z=0$ to navigate anisotropic media \cite{PRXLife.2.033004}. 

\begin{figure*}
\includegraphics[width=\textwidth]{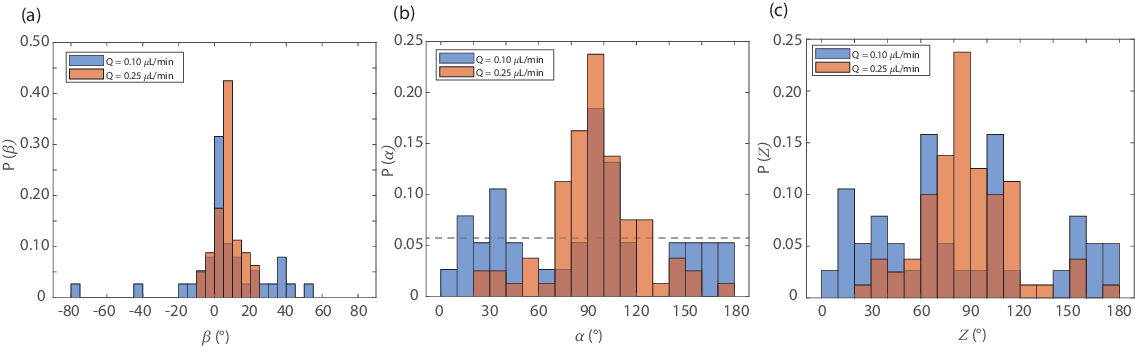}
\includegraphics[width=\textwidth]{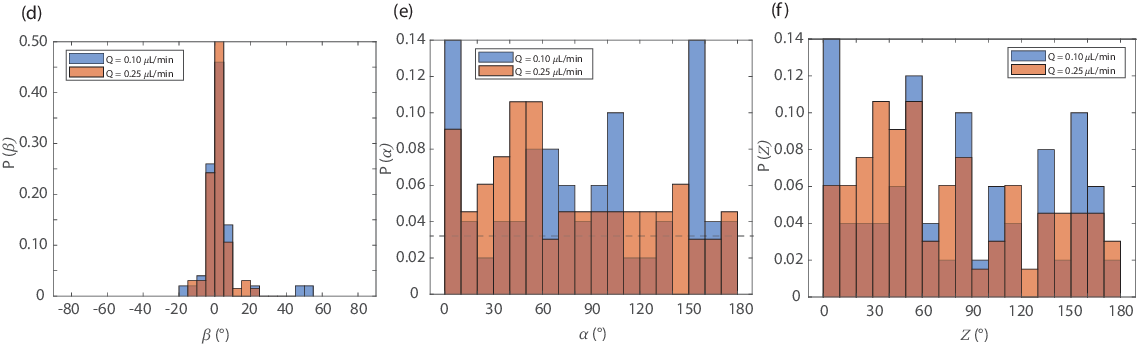}
\caption{\label{AngleHistograms} Histograms of the three angles defined in Figure \ref{DefineAngles}(a). Panels (a), (b), and (c) depict wiggling \emph{E. coli}, while (d), (e), and (f) depict non-wiggling \emph{E. coli}. (a) and (d) show distributions of trajectories $\beta$.  (b) and (e) show \emph{E. coli} orientation with respect to the trajectory, $\alpha$.  (c) and (f) show the orientation of the bacteria directors with respect to their own trajectory directions, $Z$. \added{In (a), (b) and (c) the histograms represent $N=38$ and 80 at $Q=0.10$ $\mu$L/min and $Q=0.25$ $\mu$L/min, respectively. In (d), (e) and (f) the histograms represent $N=50$ and 66 at $Q=0.10$ $\mu$L/min and $Q=0.25$ $\mu$L/min, respectively.}}
\end{figure*}



The \emph{E. coli} director orientation, $\alpha$, the trajectory angle, $\beta$, and the angle between these, $Z$, are not necessarily aligned with each other.  In $\sim83$\% of all wiggling swimmers, $\beta>0^{\circ}$, shown throughout Figure \ref{DefineAngles}. Figure \ref{DefineAngles}(b), (c) and (d) show the most common average orientations of $\alpha$ and $Z$ for wigglers in both slow and fast flows.   Approximately 30\% of all wigglers swim with both $\alpha$ and $Z<90^{\circ}$.  Another $\sim30$\% swim with both $\alpha$ and $Z>90^{\circ}$.  Another $\sim20$\% swim with $\alpha>90^{\circ}$ but $Z<90^{\circ}$. The remaining $\sim17$\% percentage of all swimmers move away from the wall, $\beta<0^{\circ}$.  However, in the slower flow, $24\%$ of wiggling \emph{E. coli} move away from the wall, and this percentage drops to $14\%$ in faster flow.  


Figure \ref{AngleHistograms} provides normalized histograms of these three angles, representing all elongated \emph{E. coli} in both flow rates, with Table II providing the average values and standard deviations.  As seen in Figure \ref{AngleHistograms}(a), $\beta$ varies widely for the lower flow rate $Q=0.10 \mu$L/min, ranging from $-75.17^{\circ}$ to $+50.52^{\circ}$. This matches the observations in Figure \ref{Trajectories},(a) in which trajectories not strongly constrained to any one direction. In faster flow, $Q=0.25 \mu$L/min, bacteria trajectories are angled toward the wall to a greater degree than in slower flow.  When $Q=0.25 \mu$L/min, $\beta$ ranges from $-9.68^{\circ}$ to $+23.95^{\circ}$.  Not only do fewer \emph{E. coli} flow toward the center, but also the range of $\beta$ is greatly narrowed.  The average $\langle\beta\rangle \sim7^{\circ}$ at both flow rates, as seen in Table II.  However the standard deviation in the faster flow is $\sim7^{\circ}$, and is more than three times that value in slower flow, $\sim23^{\circ}$.  An F-test of equality of variances confirms the statistical difference between the two populations ($p=2.77\times 10^{-7}$).  For non-wigglers, $\beta$ is narrowly distributed around $0^{\circ}$, as seen in Figure 4(d).  Non-wigglers go with the flow.

Figure \ref{AngleHistograms}(b) reveals that the swimming, wiggling \emph{E. coli} in faster flow preferentially orient nearly perpendicular to the streamlines of the flow.  The horizontal dashed line represents $\alpha=0.056$, which would be expected if the $\alpha$ values were uniformly distributed across all 18 bins.  At $Q=0.10 \mu$L/min, the distribution is not quite uniform: $\alpha$ ranges from 8.55$^{\circ}$ to 175.66$^{\circ}$, with a slight peak appearing at 84.41$^{\circ}$.  Nearly 40\% of wiggling \emph{E. coli} at $Q=0.10 \mu$L/min exhibit an orientation $80^{\circ}< \alpha <120^{\circ}$.  When the flow rate increases to $Q=0.25 \mu$L/min, the range of $\alpha$ decreases to 28.91$^{\circ}$ to 165.37$^{\circ}$.  Further, the peak around the average $\langle \alpha \rangle = 94.35^{\circ}$ becomes more prominent.  Nearly 60\% of wigglers exhibiting $^{\circ}80< \alpha <120^{\circ}$.  The behavior of the wiggling bacteria in Figure \ref{AngleHistograms}(b) contrasts sharply with the orientations of non-wiggling \emph{E. coli}.  For non-wigglers, $\alpha$ more closely resembles a uniform distribution, as seen in Figure 4(e).

\begin{figure*}
\includegraphics[width=1.07\textwidth]
{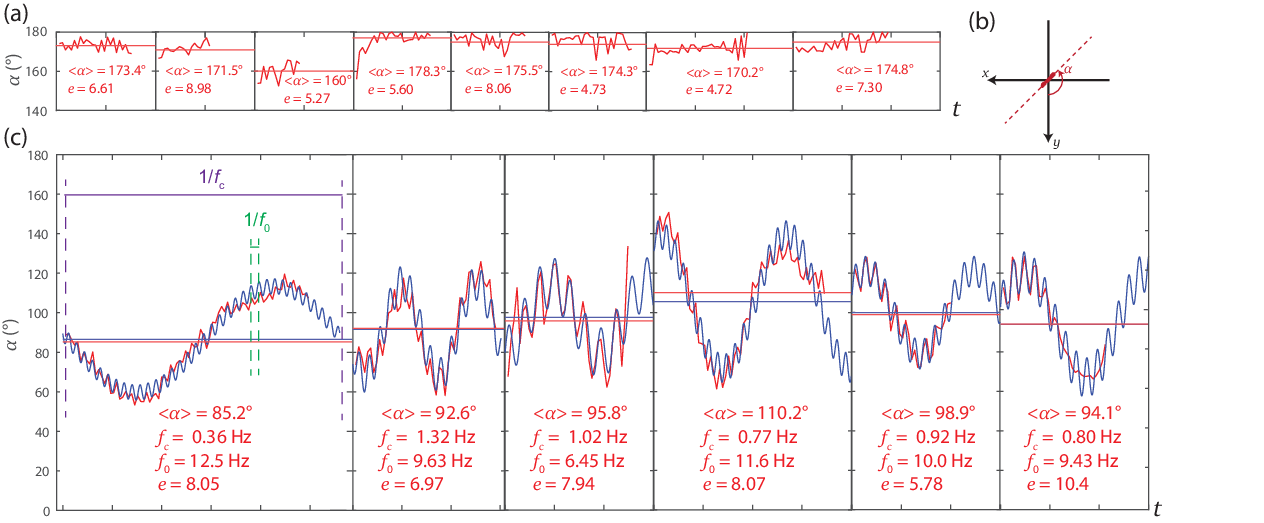}
\caption{\label{AlphaBothModes} (a) Shows eight examples of $\alpha(t)$ for non-wiggling \emph{E. coli}.   (b) Shows the definition of $\alpha$. (c) Shows six examples of $\alpha(t)$ for wiggling \emph{E. coli}, all in the faster flow, at $Q=0.25 \mu$L/min.  Measurements are shown in red, with fits to $\alpha = A_0 \sin (f_0t) + A_c \sin (f_{c}t)$ shown in blue. The horizontal lines correspond to the average $\left<\alpha\right>$ for both measurements (in red) and fit (in blue). The faster frequency $f_0$ corresponds to the \deleted{wiggling} \added{rigid body rotation} frequency identified by the procedure in Figure \ref{DefineWiggle}, while the slower frequency $f_c$ likely arises from chiral reorientation. \added{In the first example shown in (c), the vertical dashed lines indicate the period corresponding to $1/f_c$, in purple, and to $1/f_0$, in green.} On the time axis, each tick mark represents 0.5s in both (a) and (c).}
\end{figure*}


Figure \ref{AngleHistograms}(c) shows histograms of $Z$ at both flow rates \added{in the wiggling population}.  Because $\alpha$ ranges from 0-180$^{\circ}$ and $-90<\beta<90^{\circ}$, the theoretical range for $Z=\alpha-\beta$ is $-90^{\circ}<Z<270^{\circ}$.  However, we find that $Z$ ranges between 0$^{\circ}$ and 180$^{\circ}$. At the slower flow rate, there is no strongly preferred value for $Z$.  This reflects both the lack of a strongly preferred value for $\alpha$ and that the majority of $\beta$ values are small.  However, at the higher flow rate, $Z$ preferentially falls within a region around 90$^{\circ}$. An orientation $Z=90^{\circ}$ indicates an \emph{E. coli} swimming perpendicular to its own trajectory, as in the example shown in Figure \ref{DefineAngles}(a).  A t-test results in $p=7.16 \times 10^{-7}$, indicating a significant difference between the two populations, with average values of $Z=43.9^{\circ}$ and $Z=70.6^{\circ}$ at $Q=0.10$ and 0.25$\mu$L/min, respectively.  As with $\alpha$, $Z$ is nearly evenly distributed for non-wigglers, as seen in Figure 4(f).


\subsection{\label{sec:trajectories}Orientation of non-wigglers is consistent with Jeffery orbits; wigglers show chiral reorientation}


The orientation dynamics of an \emph{E. coli} director with respect to the flow suggests the absence or presence of Jeffery orbits. Rods in shear flow rotate along their long axes, as first described by Jeffery in 1922 and then expounded by Bretherton in 1962 \cite{jeffery1922motion, bretherton1962motion}.  Jeffery orbits cause rotation in the director orientation $\alpha$. The rotation period, $T$, depends on the aspect ratio of the rod $e$ and scales inversely with shear rate: 

\begin{equation}
\label{eq:jeffery}
T = \frac{2\pi}{\dot{\gamma}}\left(e+\frac{1}{e}\right)
\end{equation}

\noindent The aspect ratio of wigglers and non-wigglers alike ranges from $3.8 < e < 15.4$.  Estimating shear rate as $\dot{\gamma}\sim 2Q/H^3$, $T$ ranges from $\sim20$s, for \emph{E. coli} with the smallest $e$ in faster flow, to $>3$ minutes, for \emph{E. coli} with the largest $e$ in slower flow.  However, the residence time of the bacteria in the field of view, $\tau_r$, also scales inversely with $\dot{\gamma}$.  Therefore, the ratio $\tau_r/T$, which represents the fraction of the Jeffery orbit observable in the field of view, depends only on bacteria aspect ratio $e$.  For a short bacteria, $e=3.8$, the field of view would represent $\sim3$\% of a complete orbit, corresponding to $\sim10^{\circ}$ of rotation through the course of its trajectory.  For long bacteria, $e=15.4$, $<1$\% of a complete orbit would be viewable \added{in the microscopy video}, corresponding to $\sim3^{\circ}$ of rotation.

\begin{figure*}
\includegraphics{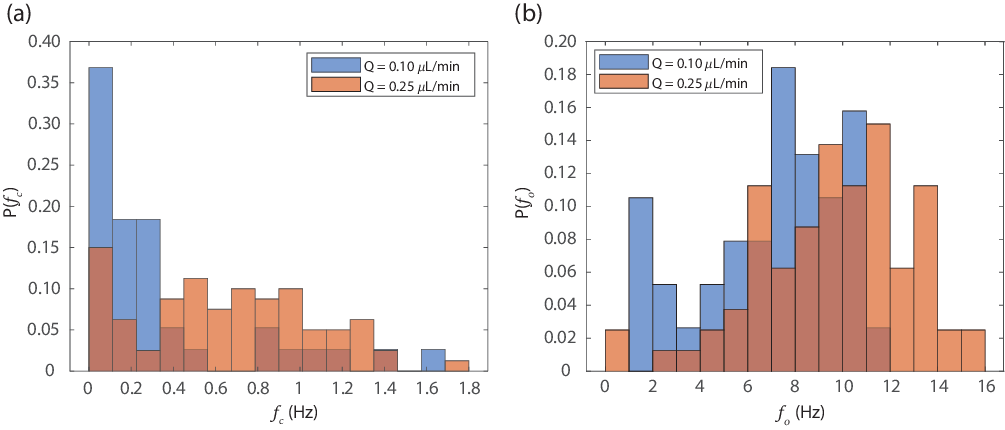}
\caption{\label{Frequency} Distributions of chiral reorientation frequency $f_c$, in (a), and rigid body rotation frequency $f_0$, in (b).  \added{Slower chiral orientation corresponds to the longer period contribution to the signals shown in Figure \ref{AlphaBothModes}c.  The faster rigid body rotation corresponds to the shorter period contribution to the signals in Figure \ref{AlphaBothModes}c.} Results for both flow rates are shown in each plot, with $Q=0.1 \mu$ L/min in blue and $Q=0.25 \mu$ L/min in orange. The distributions of $f_0$ in slow and fast flow are centered around mean values of $\langle f_0 \rangle =6.59\pm3.36$ Hz and $\langle f_0 \rangle = 9.55\pm3.17$ Hz, respectively. \added{In both panels the histograms are based on $N=38$ at $Q=0.10$ $\mu$L/min and 80 at $Q=0.25$ $\mu$L/min.} \deleted{A two sample t-test suggests the populations are distinct ($\alpha=0.05$, $p=7.4\times10^{-6}$).}}
\end{figure*}

The calculation suggesting that Jeffery orbits are much longer than the \emph{E. coli} residence time suggests two important consequences.  First, $\alpha(t)$ would remain roughly constant during the course of a single bacteria's trajectory if the bacteria were rotating in a Jeffery orbit.  Indeed, non-wigglers exhibit roughly constant $\alpha(t)$, with eight individual examples shown in Figure \ref{AlphaBothModes}(a). Each tick mark on the time axis indicates 0.5s.  Each example of non-wiggler behavior is labeled with both $\left< \alpha \right>$ and $e$; \deleted{with} the horizontal line \deleted{indicating} \added{indicates} $\left< \alpha \right>$.  While each trace of $\alpha$ is not perfectly constant through its trajectory, any fluctuations are not well fit by sine curves (as also seen in Figures \deleted{S4} \added{S7} and \deleted{S5} \added{S8}). Figure \ref{AlphaBothModes}(b) defines $\alpha$. The second consequence of a slow Jeffery orbit is that, given a roughly constant $\left< \alpha \right>$ for each \deleted{bacteria} \added{bacterium}, we expect a collection of rods to be oriented with $\left< \alpha \right>$ randomly distributed between $0^{\circ}$ and $180^{\circ}$. Indeed, this is the case for the orientation of non-wigglers, as seen in Figure \ref{AngleHistograms}(e). $P(\alpha)$ is relatively flat at both flow rates, with no strongly preferred orientation of the bacteria director.  Uniformity in both $\alpha(t)$ and in $P(\alpha)$ suggests \deleted{that} the non-wiggling bacteria trace out Jeffery orbits \added{as expected for non-motile rigid rods}.

However, the orientation of wiggling \emph{E. coli} is unlike that of non-wigglers. Wiggling \emph{E. coli} prefer to orient roughly perpendicular to streamlines, as seen in Figure \ref{AngleHistograms}(b).  Also unlike non-wigglers, $\alpha$ varies through the course of the trajectory.  Figure \ref{AlphaBothModes}(b) shows $\alpha(t)$ for a collection of six wiggling \emph{E. coli} in $Q=0.25 \mu$L/min.  Each trace corresponds to a single \emph{E. coli}, with measurements of $\alpha (t)$ shown in red.  Two distinct behaviors emerge: a large amplitude, slow sinusoidal oscillation overlaid with a smaller amplitude, higher frequency one.  The blue lines represent fits to $\alpha = A_0 \sin (f_0t+\phi_0) + A_c \sin (f_{c}t+\phi_c)$ where $f_0$ represents the faster frequency \added{of rigid body rotation} and $f_c$ the slower \deleted{one} \added{frequency. To fit $\alpha(t)$ we use $f_0$ as measured using the Fourier analysis described in Figure \ref{DefineWiggle}.  Five fitting parameters remain: the amplitudes $A_i$, phase shifts $\phi_i$, and $f_c$. The initial guess for $\phi_0$ is typically $\phi_0\approx 0$, while initial values for the other parameters are determined by visual inspection of the data traces for $\alpha(t)$.  A least-squares fitting algorithm is used to obtain final values of $A_0$, $A_c$, $\phi_0$, $\phi_c$, and $f_c$.} Each of the six examples \added{in Figure \ref{AlphaBothModes}(b)} are labeled with the corresponding frequencies and the bacteria aspect ratio $e$.  The two oscillation frequencies  are separated by approximately an order of magnitude. \added{In the first example shown, the vertical dashed lines indicate the period corresponding to $1/f_c$ in purple, and to $1/f_0$ in green.} The horizontal lines indicate $\left<\alpha\right>$ for the trajectory (in red) and for one full wavelength of the fit (in blue). 

\begin{figure*}
\includegraphics{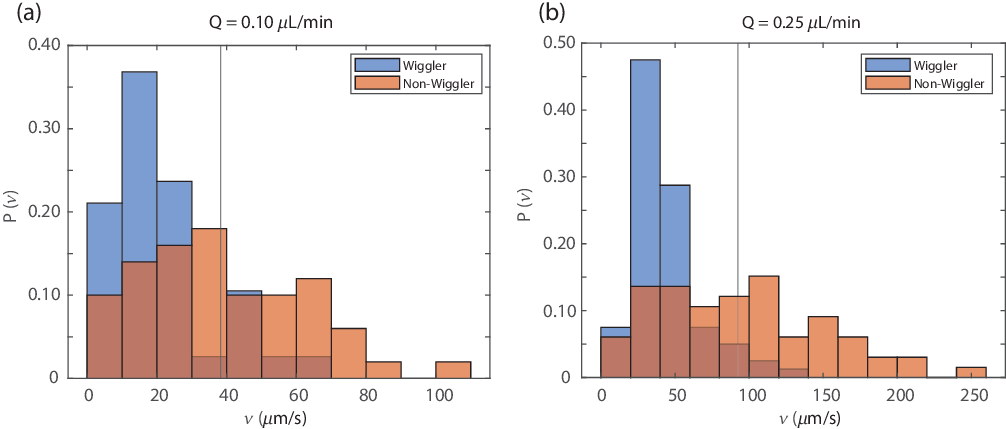}
\caption{\label{Velocity} (a) and (b) show histograms of the \emph{E. coli} velocity $v$.  At each flow rate, the distribution of velocities is significantly different when comparing wigglers to non-wigglers.  The vertical lines indicate the non-wiggler average velocity $\left< v_{\text{nw}} \right>$ at each flow rate. \added{In (a) the histograms are based on $N=38$ wigglers and 50 non-wigglers. In (b) the histograms are based on $N=80$ wigglers and 66 non-wigglers.}}
\end{figure*}

The slower frequency oscillation in $\alpha(t)$ seen in Figure \ref{AlphaBothModes}(b) may be explained by chiral reorientation.  This phenomenon appears both for \added{normal} bacteria in shear flow and for passive spheres with rigid chiral tails attached, and occurs in addition to Jeffery orbits \cite{marcos2006microorganisms, mathijssen2019oscillatory, jing2020chirality, zottl2023asymmetric}. Orientation with respect to the flow direction, $\alpha$, oscillates due to the chirality of an object in flow, with a reorientation rate $f_c = \dot{\gamma} \nu$. \deleted{, where $\nu$ is} The chiral strength of the object \added{$\nu$ accounts for hydrodynamic interactions and drag} \cite{zottl2023asymmetric, tung2015emergence, mathijssen2019oscillatory}. For spheres with rigid tails attached, both the shape and pitch of the helical tail and size and of the spherical head contribute to $\nu$.  In experiments on spheres with rigid chiral tails, in $\dot{\gamma}=30$ 1/s, oscillation is observed at a frequency $f_c \sim 1$ Hz and around an average $\sim \pm 90^{\circ}$ \cite{zottl2023asymmetric}. Damping occurs over long times.  In our elongated \emph{E. coli} experiments, shear rates are $\dot{\gamma}\sim 1$ 1/s. If the chiral strength of elongated bacteria were similar to that of the spheres with tails attached, we would expect the oscillation rate to scale with shear rate only, and thus we would expect $f_c\sim 0.03$Hz. However, in bacteria, chiral strength $\nu$ depends in a non-trivial way on shape, both of the helical flagella bundle and the cell body, and can be estimated using resistive force theory \added{to account for hydrodynamic interactions} \cite{marcos2012bacterial, mathijssen2019oscillatory, tung2015emergence}.  In general, larger cell bodies have larger $\nu$ \cite{jing2020chirality}.  Therefore, we could expect a faster reorientation rate than $f_c\sim 0.03$Hz.  In the six examples seen in Figure \ref{AlphaBothModes}c, the slower oscillation of $\alpha(t)$ ranges from $f_c = 0.36$ to 1.32 Hz.  We observe no strong dependence of $f_c$ on $e$.  However, within the entire population of bacteria, $e$ is roughly Gaussian distributed, with a standard deviation of only $\sim25$\% (Figure S1).  Taken together, these observations suggest that $f_c$ is reasonably described as arising from chiral reorientation.

The full distributions of chiral reorientation frequency for all wiggling \emph{E. coli} are shown in Figure \ref{Frequency}(a), in both flow rates.  Interestingly, the overall range of $f_c$ is roughly similar in the two flow rates, ranging from $\sim 0.007$ to $\sim 1.7$ or 1.8 Hz.  However, the distribution of $f_c$ in slower flow has a tail at higher frequencies, while $f_c$ in faster flow seems to be more normally distributed.  The average chiral reorientation rate shifts, nearly with $\dot{\gamma}$.  In slower flow, $\left< f_c \right> = 0.34$ Hz.  In flow sped up by a factor of 2.5, chiral reorientation speeds up by a factor of 1.9, to $\left< f_c \right> = 0.65$ Hz.  A t-test suggests a statistically significant difference between the two populations shown in Figure \ref{Frequency}(a), with $p=2.3\times10^{-4}$.



In the dynamics of $\alpha(t)$ in Figure \ref{AlphaBothModes}(b), the high frequency oscillation, $f_0$, corresponds to \deleted{the frequency with which bacteria curvature $a$ changes, as described above in the definition of wiggling.  However, while a Fourier transform is needed to extract $f_0$ from} \added{the wiggling defined} \deleted{$a(t)$, as} in Figure \ref{DefineWiggle}. \deleted{the behavior of $\alpha(t)$ reveals that $f_0$ is indeed associated with sinusoidal behavior.  This observation suggests that $f_0$ corresponds to rigid body rotation.}  In quiescence, rigid body rotation appears alone, and manifests as a sinusoidal oscillation in bacteria curvature $a$ (Figure \deleted{S2} \added{S4}).  In flow, \deleted{however,} rigid body rotation is joined by \deleted{the additional phenomenon of} chiral reorientation, and thus the curvature of the bacteria no longer appears to oscillate as a simple sine (Figure \ref{DefineWiggle}).  The superposition of these two phenomena \deleted{therefore explains why} \added{causes} rigid body rotation \added{to} manifest differently in flow than in quiescence.  

In comparing the two flow rates, the wiggling rigid body rotation appears to speed up \added{slightly} in faster flow, as shown in the distributions of $f_0$ \deleted{seen} in Figure \ref{Frequency}(b).  The lower bound on $f_0$ does not change significantly with $Q$.  In both flow rates, the slowest rotations are $f_0\sim 1$ Hz, and of the same order of magnitude as the shear rate.  However, the upper bound on $f_0$ increases in faster flow. In slower flow, $f_0$ ranges up to $12$Hz.  In faster flow, the distribution appears to be more Gaussian, with $f_0$ ranging up to $16$Hz.  However, while $Q$ increases by a factor of 2.5, the average $\langle f_0 \rangle$ increases by $\sim50\%$, from 6.5 to 10 Hz. A two-factor t-test confirms the significant difference between the two populations \deleted{of $f_0$} ($p=8.5\times10^{-7}$).

\begin{table*}
    \centering
    \begin{tabular}{|c|c|c|c|c|} \hline 
         \textbf{Category} & 
         \makecell{\boldmath {$f_0\sim O(10)$} \textbf{Hz} \\ \textbf{oscillation in $\alpha$}} & \makecell{\textbf{Comparing} \\ \textbf{low and high $\dot{\gamma}$}}  & \makecell{\boldmath {$f_0\sim O(1)$} \textbf{Hz}  \\ \textbf{oscillation in $\alpha$}} & \makecell{\textbf{Comparing} \\ \textbf{low and high $\dot{\gamma}$}}  \\ \hline 
          \makecell{\textbf{Motile} \\ \textbf{wigglers}}
&   \makecell{Rigid body rotation: \\ propulsion by \\ rotating flagellar bundle}
 & \makecell{$f_0$ increases 50\% as $\dot{\gamma}$  \\ increases by factor of 2.5} & \makecell{Chiral reorientation \\ due to flagellar chirality}
& \makecell{$f_c$ increases by factor of 1.9 \\ as $\dot{\gamma}$ increases by factor of 2.5}  \\ \hline 
         \makecell{\textbf{Non-motile} \\ \textbf{non-wigglers}} &  \makecell{No rigid body rotation: \\ no rotating flagellar bundle}
 & N/A & \makecell{No chiral reorientation: \\ even distribution of $\alpha$ \\ suggests Jeffery orbits only }
& \makecell{Jeffery orbit period \\ longer than residence time \\ $T \gg \tau_r$ at both $\dot{\gamma}$ } \\ \hline 
    \end{tabular}
    \caption{\added{Summary of bacteria orientation $\alpha$ with mechanistic descriptions of the measured oscillation frequencies.}}
    \label{tab:summary}
\end{table*}

In addition to clearly illustrating differences in the dynamics of $\alpha(t)$ Figure \ref{AlphaBothModes} also hints at a difference in velocities.  That is, all trajectories of $\alpha(t)$ shown in \added{both} Figure \ref{AlphaBothModes}(a) and (c) correspond to \emph{E. coli} passing through the entire \added{length of the} field of view, but represent different elapsed times.  \added{Non-motile,} non-wiggling \emph{E. coli} trajectories \added{trace the flow and} are faster than wiggling trajectories by as much as a factor of 3.  The rigid body rotation of the wiggling \emph{E. coli}, combined with chiral reorientation, slows them down with respect to the fluid. Figure \ref{Velocity} shows histograms of $\langle v\rangle$, for wiggling and non-wiggling \emph{E. coli}, at $Q=0.10 \mu$L/min, in (a), and at $Q=0.25 \mu$L\added{/}min, in (b).  Average values and standard deviations are shown in Table II. At both flow rates, \added{the motile wigglers are slower than the non-motile non-wigglers which trace the fluid flow.} \deleted{non-wiggling \emph{E. coli}, tracing the flow, flow faster than wigglers.}  In the slower flow, wigglers have $\langle v\rangle = 21.3 \pm 14.6 \mu$m/s, while for non-wigglers, $\langle v\rangle = 38.8 \pm 23.9 \mu$m/s ($p=1.43 \times 10^{-4}$).  On average, \deleted{non wigglers are $82\%$ faster than wigglers} \added{wigglers are 45\% slower than non-wiggler tracerss when $Q=0.10$ $\mu$L/min}.  In faster flow, wigglers have $\langle v\rangle = 43.4 \pm 22.0 \mu$m/s, while for the non-wigglers, $\langle v\rangle = 93.5 \pm 55.6 \mu$m/s ($p=1.09 \times 10^{-11}$). Again, on average, \deleted{non-wigglers are faster than wigglers, this time by more than $100\%$} \added{wigglers are 53\% slower than non-wiggler tracers}.  The vertical lines in Fig. \ref{Velocity} indicate the average non-wiggler velocity, $\left< v_{\text{nw}} \right>$, at each flow rate.  
\deleted{The wiggling rigid body rotation} \added{The chiral reorientation of elongated, motile bacteria} appears to slow \deleted{the}\added{their} overall velocity \deleted{of the motile bacteria} to a \added{somewhat} greater degree in the faster flow. \added{Preferential orientation of the elongated cell bodies nearly perpendicular to both streamlines and trajectories subjects these motile bacteria to increased drag, thereby slowing them with respect to the fluid velocity.  Figure \deleted{S8} \added{S14} plots the \% decrease in velocity of the motile, wiggling bacteria with respect to the flow field as a function of the average position of each motile wiggler trajectory across the channel in $x$.}

While the motility of elongated \emph{E. coli} in flow can, in general, be described as a combination of rigid body rotation and chiral reorientation, in several instances we observe \emph{E. coli} to behave in unexpected ways.  Some \emph{E. coli} are observed to wiggle and then stop wiggling, or rotating, during the course of their trajectory. The final example in Figure \ref{AlphaBothModes}(b) shows this: the high frequency wiggling stops in the middle of the trajectory, at approximately 0.5s.  The end of rigid body rotation may suggest that the flagellar bundle has stopped rotating.  Interestingly, despite the end of rigid body rotation, the slower chiral reorientation continues for another half second.  Some \emph{E. coli} observed to swim upstream at the wall can also be seen to momentarily pause wiggling as they fall back into the flow \added{as shown in Figure S9 and S10, each with a corresponding video online.} \deleted{(SI videos Upstream1 and Upstream2).} 




In another example unlike the majority, a wiggling \emph{E. coli} swims away from the wall in $Q=0.10 \mu$L/min, with $\beta\sim-80^{\circ}$. In the trajectory of this bacteria, seen in \added{Figure S12 (corresponding video online)} \deleted{an SI video (WallAway)}, it first swims upstream along the wall and then continues swimming upstream as it moves toward the channel center.  This \emph{E. coli} swims with a relatively constant velocity as it moves away from the wall, as shown in Figure \deleted{S9} \added{S15}.  The roughly constant velocity indicates this observation is unrelated to shear-induced drift away from a wall occurring in emulsions and suspensions of soft colloids \cite{chan1979motion, kumar2014flow,marnoto2023application}.  Rather, the unique behavior of this swimmer may be in part explained by rheotaxis.  In the initial trajectory of this bacteria away from the wall, it swims toward its own left side as it moves toward the channel center.




\section{\label{sec:conc}Conclusion}




Theoretical predictions of rheotaxis suggest that motile, chiral, rod-like bacteria in external flow swim downstream and to the left, in the vorticity direction.  Chiral reorientation \deleted{also} leads to a reorientation of the bacteria director nearly perpendicular to the flow direction.  These predictions have been observed experimentally in normally-shaped \emph{E. coli}, which are short rods.  We provide measurements of these phenomena in significantly longer, filamentous, swimming \emph{E. coli} for the first time. \deleted{, with additional observations.  We characterize the wiggling motion of stressed, filamentous \emph{E. coli} in flow through a microchannel at two different external flow rates.  We extract several metrics of swimming trajectories, including velocity and the evolution of bacteria shape and orientation, both of which change throughout a single trajectory.  The} 

Filamentous \emph{E. coli} swim toward the wall, as predicted by rheotaxis, \added{but with meandering paths at a slower flow rate and straighter ones over a narrower range of trajectory angles} \deleted{but do so far more reliably} at a faster flow rate. \deleted{Consequently, the extended length of these bacteria reveals that} Chiral reorientation causes their bodies to align \deleted{, not perpendicular to the flow direction, but rather} perpendicular to the direction of the bacteria's own trajectory \added{subjecting them to increased drag}.  \deleted{Further,} The oscillatory behavior of chiral reorientation is coupled with a higher frequency oscillation consistent with rigid body rotation \added{of the elongated, buckled rods}.  \deleted{Rigid body rotation arises due to the segmentation of the filamentous bacteria, giving these \emph{E. coli} swimming in external flows the appearance of wiggling.} 
\deleted{Interestingly,} Approximately half of the filamentous \emph{E. coli} in flow exhibit none of the behaviors just described. \deleted{in pressure driven flow.} \added{These non-motile bacteria} \deleted{Instead, they} flow along streamlines, with no preferred orientation of their directors, \deleted{.  Their orientations are} consistent with Jeffery orbits.  \deleted{These} Non-wigglers \added{are passive, achiral rods that} trace the flow, moving faster than the \deleted{wiggling} \added{motile} population.  \deleted{ That is: non-wiggling \emph{E. coli} behave like passive, achiral rods.}   \deleted{Unbundled flagella would explain the lack} \added{The absence} of rigid body rotation \added{and chiral reorientation could be explained by unbundled flagella, damaged flagella, or cell death.} \deleted{in this population.  Without a flagellar bundle, chirality is also absent, explaining the lack of chiral reorientation.  These non-wiggling \emph{E. coli} do not exhibit the hallmarks of motile \emph{E. coli}, and may be dead.}  \added{Table \ref{tab:summary} provides a summary.  Rigid body rotation and chiral reorientation require a chiral bundle of rotating flagella, in other words: motility.  Non-motile non-wigglers, without a rotating chiral flagellar bundle, exhibit neither rigid body rotation nor chiral reorientation.}



Our results open up several directions for future study \added{including theoretical approaches.  Chiral reorientation dynamics in short rods can be described by rotation in shear and parabolic flow.  In this framework, hydrodynamic interactions are accounted for using a single chiral strength parameter to describe the object \cite{jing2020chirality, mathijssen2019oscillatory, tung2015emergence}.  A full theoretical description of elongated, chiral, rotating swimmers will likely require descriptions of rigid body rotation and chiral reorientation combined with hydrodynamic drag on elongated bodies and hydrodynamic interactions with the wall. We anticipate the possibility of a critical shear-induced transition in the trajectories of elongated bacteria from the widely spread histogram of $\beta$ at the lower flow rate to the much narrower distribution of trajectory angles at the higher flow rate.  Such a transition could indicate flow conditions in which stressed bacteria preferentially reach walls in shear rates relevant to clinical settings. Attachment may be  determined by a competition between a minimum shear rate needed to enhance flow toward the wall and an upper shear rate beyond which bacteria will be swept away from it.}



Because filamentation can be induced by sub-MIC antibiotic treatments, \deleted{these} greatly elongated swimmers provide important insights into the behavior of bacteria that are resistant to drug treatments.  \deleted{Thus, it is important to both measure and predict swimming behavior as a function of bacteria body length over a broad range.}  \deleted{The degree to which filamentous bacteria swim toward side walls, as a function of both body length and external shear rate, could indicate how easily they can reach and then attach to surfaces.} \added{A failure of antibiotics to kill bacteria may instead encourage bacteria to reach surfaces in flow more easily in certain flow conditions. Understanding how motility reduction impacts swimming in microchannel flows, and especially how it facilitates wall attachment, could suggest design rules for flow conditions that may prevent biofilm formation in clinical tubing. Flagellar tags and live-dead assays coupled with rigid body rotation and chiral reorientation measurements could reveal how stages like flagellar debundling and rotation cessation both determine swimming behavior and precede death as a function of antibiotic dosage. Considering even lower antibiotic doses could lend insight into the behavior of stressed bacteria swimming in the earth's subsurface pores, as groundwater near wastewater treatment plants may be contaminated by antibiotics.}

\section{Acknowledgements}
YA and JEH gratefully acknowledge funding from the U.S. Department of Agriculture CSREES HSI (No. 2006-38422-17086). The \emph{E. coli} used in this study were obtained from the Yale \emph{E. coli} Genetic Stock Center. The authors gratefully acknowledge helpful conversations with Tom Powers, Arnold Mathijssen, Jacinta Conrad and Jay Tang.

\section{Conflict of Interest}
The authors have no conflicts to disclose.

\section{Data Availability Statement}
Data are available upon reasonable request to the corresponding author.



\section{Supplemental Material}
See Supplemental Material at [URL will be inserted by publisher] for a histogram of \emph{E. coli} lengths, multimedia descriptions, an example of sinusoidal dynamics in quiescent swimming, supporting documentation for classifying \emph{E. coli} as ``wiggling'' or ``non-wiggling,'' example dynamics and trajectories of non-motile non-wigglers, histograms of path tortuosity, velocity profiles of the wigglers with the flow subtracted, and an example of constant-velocity swimming away from the wall.



\begin{thebibliography}{10}

\bibitem{aranson2022bacterial}
Igor~S Aranson.
\newblock Bacterial active matter.
\newblock {\em Reports on Progress in Physics}, 85(7):076601, 2022.

\bibitem{koch2011collective}
Donald~L Koch and Ganesh Subramanian.
\newblock Collective hydrodynamics of swimming microorganisms: living fluids.
\newblock {\em Annual Review of Fluid Mechanics}, 43(1):637--659, 2011.

\bibitem{zhang2010collective}
He-Peng Zhang, Avraham Be’er, E-L Florin, and Harry~L Swinney.
\newblock Collective motion and density fluctuations in bacterial colonies.
\newblock {\em Proceedings of the National Academy of Sciences}, 107(31):13626--13630, 2010.

\bibitem{costerton1999bacterial}
J~William Costerton, Philip~S Stewart, and E~Peter Greenberg.
\newblock Bacterial biofilms: a common cause of persistent infections.
\newblock {\em Science}, 284(5418):1318--1322, 1999.

\bibitem{donlan2002biofilms}
Rodney~M Donlan and J~William Costerton.
\newblock Biofilms: survival mechanisms of clinically relevant microorganisms.
\newblock {\em Clinical microbiology reviews}, 15(2):167--193, 2002.

\bibitem{sauer_biofilm_2022}
Karin Sauer, Paul Stoodley, Darla~M. Goeres, Luanne Hall-Stoodley, Mette Burmølle, Philip~S. Stewart, and Thomas Bjarnsholt.
\newblock The biofilm life cycle: expanding the conceptual model of biofilm formation.
\newblock {\em Nature Reviews Microbiology}, 20(10):608--620.
\newblock Number: 10 Publisher: Nature Publishing Group.

\bibitem{lebeaux_vitro_2013}
David Lebeaux, Ashwini Chauhan, Olaya Rendueles, and Christophe Beloin.
\newblock From in vitro to in vivo models of bacterial biofilm-related infections.
\newblock {\em Pathogens}, 2(2):288--356, 2013.

\bibitem{verstraeten2008living}
Natalie Verstraeten, Kristien Braeken, Bachaspatimayum Debkumari, Maarten Fauvart, Jan Fransaer, Jan Vermant, and Jan Michiels.
\newblock Living on a surface: swarming and biofilm formation.
\newblock {\em Trends in microbiology}, 16(10):496--506, 2008.

\bibitem{yazdi2012bacterial}
Shahrzad Yazdi and Arezoo~M Ardekani.
\newblock Bacterial aggregation and biofilm formation in a vortical flow.
\newblock {\em Biomicrofluidics}, 6(4), 2012.

\bibitem{carrillo2025preventing}
Juan Pablo Carrillo-Mora, Moniellen Pires Monteiro, V.~I. Marconi, Mar{\'\i}a Luisa Cordero, Ricardo Brito and Rodrigo Soto.
\newblock Preventing clustering of active particles in microchannels.
\newblock {\em Communications Physics}, 8(1):374, 2025.

\bibitem{rusconi2014bacterial}
Roberto Rusconi, Jeffrey~S. Guasto and Roman Stocker.
\newblock Bacterial transport suppressed by fluid shear.
\newblock {\em Nature physics}, 10(3):212--217, 2014.

\bibitem{figueroa2015living}
Nuris Figueroa-Morales, Gast{\'o}n~Leonardo Mi{\~n}o, Aramis Rivera, Rogelio Caballero, Eric Cl{\'e}ment, Ernesto Altshuler, and Anke Lindner.
\newblock Living on the edge: transfer and traffic of e. coli in a confined flow.
\newblock {\em Soft matter}, 11(31):6284--6293, 2015.

\bibitem{ajulo2024global}
Samuel Ajulo and Babafela Awosile.
\newblock Global antimicrobial resistance and use surveillance system (glass 2022): Investigating the relationship between antimicrobial resistance and antimicrobial consumption data across the participating countries.
\newblock {\em PLoS One}, 19(2):e0297921, 2024.

\bibitem{ventola_antibiotic_2015}
C.~Lee Ventola.
\newblock The antibiotic resistance crisis.
\newblock {\em Pharmacy and Therapeutics}, 40(4):277--283.

\bibitem{canton_emergence_2011}
Rafael Cantón and María-Isabel Morosini.
\newblock Emergence and spread of antibiotic resistance following exposure to antibiotics.
\newblock {\em {FEMS} microbiology reviews}, 35(5):977--991.

\bibitem{tang2023antimicrobial}
Ka~Wah~Kelly Tang, Beverley~C Millar, and John~E Moore.
\newblock Antimicrobial resistance (amr).
\newblock {\em British journal of biomedical science}, 80:11387, 2023.

\bibitem{khan2022filamentous}
Fazlurrahman Khan, Geum-Jae Jeong, Nazia Tabassum, Akanksha Mishra, and Young-Mog Kim.
\newblock Filamentous morphology of bacterial pathogens: regulatory factors and control strategies.
\newblock {\em Applied Microbiology and Biotechnology}, 106(18):5835--5862, 2022.

\bibitem{zhang_PNAS2024_filament}
Dongxue Zhang, Fan Yin, Qin Qin, and Liang Qiao.
\newblock Molecular responses during bacterial filamentation reveal inhibition methods of drug-resistant bacteria.
\newblock {\em Proceedings of the National Academy of Sciences}, 120(27):e2301170120, 2023.

\bibitem{wehrens_size_2018}
Martijn Wehrens, Dmitry Ershov, Rutger Rozendaal, Noreen Walker, Daniel Schultz, Roy Kishony, Petra~Anne Levin, and Sander~J. Tans.
\newblock Size laws and division ring dynamics in filamentous \textit{Escherichia coli} cells.
\newblock {\em Current Biology}, 28(6):972--979.e5.

\bibitem{phan2018emergence}
Trung V Phan, Ryan J Morris, Ho Tat Lam,  Phuson Hulamm, Matthew E Black, Julia  Bos, and Robert H Austin.
\newblock Emergence of Escherichia coli critically buckled motile helices under stress.
\newblock {\em Proceedings of the National Academy of Sciences}, 115(51):12979--12984, 2018.

\bibitem{nadal2025coli}
Marta Nadal, L{\'e}na Guitou, Iago Diez,  Juan Hurtado, Alejandro Mart{\'\i}nez, Iago Grobas, and Javier Buceta.
\newblock E. coli filament buckling modulates Min patterning and cell division.
\newblock {\em Nature Communications}, 16(1):8193, 2025. 

\bibitem{jeffery1922motion}
George~Barker Jeffery.
\newblock The motion of ellipsoidal particles immersed in a viscous fluid.
\newblock {\em Proceedings of the Royal Society of London. Series A}, 102(715):161--179, 1922.

\bibitem{bretherton1962motion}
Francis~P Bretherton.
\newblock The motion of rigid particles in a shear flow at low reynolds number.
\newblock {\em Journal of Fluid Mechanics}, 14(2):284--304, 1962.

\bibitem{zottl2023asymmetric}
Andreas Z{\"o}ttl, Francesca Tesser, Daiki Matsunaga, Justine Laurent, Olivia Du~Roure, and Anke Lindner.
\newblock Asymmetric bistability of chiral particle orientation in viscous shear flows.
\newblock {\em Proceedings of the National Academy of Sciences}, 120(45):e2310939120, 2023.

\bibitem{berg.1999}
Richard~M Berry and Howard~C Berg.
\newblock Torque generated by the flagellar motor of escherichia coli while driven backward.
\newblock {\em Biophysical Journal}, 76(1):580--587, 1999.

\bibitem{lowe1987rapid}
Graeme Lowe, Markus Meister, and Howard~C Berg.
\newblock Rapid rotation of flagellar bundles in swimming bacteria.
\newblock {\em Nature}, 325(6105):637--640, 1987.

\bibitem{wadhwa.2021}
Navish Wadhwa, Yuhai Tu, and Howard~C Berg.
\newblock Mechanosensitive remodeling of the bacterial flagellar motor is independent of direction of rotation.
\newblock {\em Proceedings of the National Academy of Sciences}, 118(15):e2024608118, 2021.

\bibitem{macnab1977bacterial}
Macnab, Robert M.
\newblock Bacterial flagella rotating in bundles: a study in helical geometry.
\newblock {\em Proceedings of the National Academy of Sciences}, 74(1):221--225, 1977.

\bibitem{powers2002role}
Powers, Thomas R.
\newblock Role of body rotation in bacterial flagellar bundling.
\newblock {\em Physical Review E}, 65(4):040903, 2002.



\bibitem{PRXLife.2.033004}
Ameya~G. Prabhune, Andy~S. Garc\'{\i}a-Gordillo, Igor~S. Aranson, Thomas~R. Powers, and Nuris Figueroa-Morales.
\newblock Bacteria navigate anisotropic media using a flagellar tug-of-oars.
\newblock {\em PRX Life}, 2:033004, Jul 2024.

\bibitem{LAUGA2006400}
Eric Lauga, Willow~R. DiLuzio, George~M. Whitesides, and Howard~A. Stone.
\newblock Swimming in circles: Motion of bacteria near solid boundaries.
\newblock {\em Biophysical Journal}, 90(2):400--412, 2006.

\bibitem{lauga_bacterial_2016}
Eric Lauga.
\newblock Bacterial hydrodynamics.
\newblock {\em Annual Review of Fluid Mechanics}, 48(1):105--130, 2016.

\bibitem{PhysRevX.7.011010}
Silvio Bianchi, Filippo Saglimbeni, and Roberto Di~Leonardo.
\newblock Holographic imaging reveals the mechanism of wall entrapment in swimming bacteria.
\newblock {\em Phys. Rev. X}, 7:011010, 2017.

\bibitem{diluzio2005escherichia}
Willow~R DiLuzio, Linda Turner, Michael Mayer, Piotr Garstecki, Douglas~B Weibel, Howard~C Berg, and George~M Whitesides.
\newblock Escherichia coli swim on the right-hand side.
\newblock {\em Nature}, 435(7046):1271--1274, 2005.

\bibitem{marcos2012bacterial}
Marcos, Henry~C Fu, Thomas~R Powers, and Roman Stocker.
\newblock Bacterial rheotaxis.
\newblock {\em Proceedings of the National Academy of Sciences}, 109(13):4780--4785, 2012.

\bibitem{mathijssen2019oscillatory}
Arnold~JTM Mathijssen, Nuris Figueroa-Morales, Gaspard Junot, {\'E}ric Cl{\'e}ment, Anke Lindner, and Andreas Z{\"o}ttl.
\newblock Oscillatory surface rheotaxis of swimming e. coli bacteria.
\newblock {\em Nature communications}, 10(1):3434, 2019.

\bibitem{hill_hydrodynamic_2007}
Jane Hill, Ozge Kalkanci, Jonathan~L. {McMurry}, and Hur Koser.
\newblock Hydrodynamic surface interactions enable escherichia coli to seek efficient routes to swim upstream.
\newblock {\em Physical Review Letters}, 98(6):068101, 2007.

\bibitem{kaya_direct_2012}
Tolga Kaya and Hur Koser.
\newblock Direct upstream motility in escherichia coli.
\newblock {\em Biophysical Journal}, 102(7):1514--1523.

\bibitem{figueroa-morales_e_2020}
Nuris Figueroa-Morales, Aramis Rivera, Rodrigo Soto, Anke Lindner, Ernesto Altshuler, and Éric Clément.
\newblock E. coli “super-contaminates” narrow ducts fostered by broad run-time distribution.
\newblock {\em Science Advances}, 6(11):eaay0155, 2020.

\bibitem{zhou2024ai}
Tingtao Zhou, Xuan Wan, Daniel~Zhengyu Huang, Zongyi Li, Zhiwei Peng, Anima Anandkumar, John~F Brady, Paul~W Sternberg, and Chiara Daraio.
\newblock AI-aided geometric design of anti-infection catheters.
\newblock {\em Science Advances}, 10(1):eadj1741, 2024.

\bibitem{jing2020chirality}
Guangyin Jing, Andreas Z{\"o}ttl, {\'E}ric Cl{\'e}ment, and Anke Lindner.
\newblock Chirality-induced bacterial rheotaxis in bulk shear flows.
\newblock {\em Science advances}, 6(28):eabb2012, 2020.

\bibitem{zhang2022enterobacter}
Zhiyu Zhang, Haoming Liu, Hamid Karani, Jon Mallen, Weijie Chen, Arpan De, Sridhar Mani, and Jay~X Tang.
\newblock Enterobacter sp. strain sm1\_hs2b manifests transient elongation and swimming motility in liquid medium.
\newblock {\em Microbiology Spectrum}, 10(3):e02078--21, 2022.

\bibitem{maki2000motility}
Nazli Maki, Jason~E Gestwicki, Ellen~M Lake, Laura~L Kiessling, and Julius Adler.
\newblock Motility and chemotaxis of filamentous cells of escherichia coli.
\newblock {\em Journal of Bacteriology}, 182(15):4337--4342, 2000.

\bibitem{duffy1998rapid}
David~C Duffy, J~Cooper McDonald, Olivier~JA Schueller, and George~M Whitesides.
\newblock Rapid prototyping of microfluidic systems in poly (dimethylsiloxane).
\newblock {\em Analytical chemistry}, 70(23):4974--4984, 1998.

\bibitem{mcdonald2000fabrication}
J~Cooper McDonald, David~C Duffy, Janelle~R Anderson, Daniel~T Chiu, Hongkai Wu, Olivier~JA Schueller, and George~M Whitesides.
\newblock Fabrication of microfluidic systems in poly (dimethylsiloxane).
\newblock {\em ELECTROPHORESIS: An International Journal}, 21(1):27--40, 2000.

\bibitem{skryabina2006disposable}
Skryabina, Elena A and Dunn, Teresa S.
\newblock Disposable infusion pumps.
\newblock {\em American Journal of Health-System Pharmacy}, 63(13):1260--1268, 2006.

\bibitem{breland2010continuous}
Breland, Burnis D.
\newblock Continuous quality improvement using intelligent infusion pump data analysis.
\newblock {\em American Journal of Health-System Pharmacy}, 67(17):1446--1455, 2010.

\bibitem{crocker1996methods}
John~C Crocker and David~G Grier.
\newblock Methods of digital video microscopy for colloidal studies.
\newblock {\em Journal of colloid and interface science}, 179(1):298--310, 1996.

\bibitem{mears.2014}
Patrick~J Mears, Santosh Koirala, Chris~V Rao, Ido Golding, and Yann~R Chemla.
\newblock \textit{Escherichia coli} swimming is robust against variations in flagellar number.
\newblock {\em eLife}, 3:e01916, 2014.

\bibitem{lutkenhaus1997bacterial}
Joe Lutkenhaus.
\newblock Bacterial cytokinesis: let the light shine in.
\newblock {\em Current Biology}, 7(9):R573--R575, 1997.

\bibitem{ebersbach2008novel}
Gitte Ebersbach, Elisa Galli, Jakob M{\o}ller-Jensen, Jan L{\"o}we, and Kenn Gerdes.
\newblock Novel coiled-coil cell division factor zapb stimulates z ring assembly and cell division.
\newblock {\em Molecular microbiology}, 68(3):720--735, 2008.

\bibitem{turner2012growth}
Linda Turner, Alan S Stern and Howard C Berg.
\newblock Growth of flagellar filaments of Escherichia coli is independent of filament length.
\newblock {\em Journal of bacteriology}, 194(10):2437--2442, 2012.

\bibitem{carrillo2025damage}
Juan Pablo Carrillo-Mora, Moniellen Pires Monteiro, An{\'\i}bal R Lodeiro, V~I Marconi and Mar{\'\i}a Luisa Cordero.
\newblock Damage and recovery of flagella in soil bacteria exposed to shear within long microchannels.
\newblock {\em Physics of Fluids}, 37(1), 2025.


\bibitem{liu2018morphological}
Yanan Liu, Brato Chakrabarti, David Saintillan, Anke Lindner, and Olivia du Roure.
\newblock Morphological transitions of elastic filaments in shear flow.
\newblock {\em Proceedings of the National Academy of Sciences}, 115(38):9438--9443, 2018.

\bibitem{amir_bending_2014}
Ariel Amir, Farinaz Babaeipour, Dustin~B. {McIntosh}, David~R. Nelson, and Suckjoon Jun.
\newblock Bending forces plastically deform growing bacterial cell walls.
\newblock {\em Proceedings of the National Academy of Sciences}, 111(16):5778--5783, 2014.

\bibitem{deng2011prl}
Yi Deng, Mingzhai Sun, and Joshua W. Shaevitz.
\newblock Direct Measurement of Cell Wall Stress Stiffening and Turgor Pressure in Live Bacterial Cells.
\newblock {\em Physical Review Letters}, 107(15):158101, 2011.

\bibitem{benhamou_how_2004}
Simon Benhamou.
\newblock How to reliably estimate the tortuosity of an animal's path: straightness, sinuosity, or fractal dimension?
\newblock {\em Journal of Theoretical Biology}, 229(2):209--220, 2004.

\bibitem{marcos2006microorganisms}
Marcos and Roman Stocker.
\newblock Microorganisms in vortices: a microfluidic setup.
\newblock {\em Limnology and Oceanography: Methods}, 4(10):392--398, 2006.

\bibitem{tung2015emergence}
Chih-kuan Tung, Florencia Ardon, Anubhab Roy, Donald~L Koch, Susan~S Suarez and Mingming Wu.
\newblock Emergence of upstream swimming via a hydrodynamic transition.
\newblock {\em Physical review letters}, 114(10):108102, 2015.

\bibitem{chan1979motion}
PC-H Chan and LG0402 Leal.
\newblock The motion of a deformable drop in a second-order fluid.
\newblock {\em Journal of fluid mechanics}, 92(1):131--170, 1979.

\bibitem{kumar2014flow}
Amit Kumar, Rafael G~Henr{\'\i}quez Rivera, and Michael~D Graham.
\newblock Flow-induced segregation in confined multicomponent suspensions: effects of particle size and rigidity.
\newblock {\em Journal of Fluid Mechanics}, 738:423--462, 2014.

\bibitem{marnoto2023application}
Sabrina Marnoto and Sara~M Hashmi.
\newblock Application of droplet migration scaling behavior to microchannel flow measurements.
\newblock {\em Soft Matter}, 19(3):565--573, 2023.

 

\end{thebibliography}


\end{document}